\documentclass[fleqn,11pt,twoside]{article}
\usepackage{amsthm,amsthm,amssymb, color, epsfig, graphics, subfigure,}
\usepackage{tikz}
\usepackage{amsmath}%[fleqn,intlimits]
\usepackage{graphicx}
\usepackage{amsmath}%[fleqn,intlimits]
\usepackage{graphicx}
\usepackage{latexsym}
\usepackage[all]{xy}
\newtheorem{proposition}{Proposition}

\makeatletter
\newcommand{\copyrightnote}[2]{{\renewcommand{\thefootnote}{}
 \footnotetext{\small\it
\begin{flushleft}
 \copyright \ #1  #2
\end{flushleft}}}}

\newcommand{\Name}[1]{\begin{flushleft}
                       \LARGE \bf #1
                      \end{flushleft}\vspace{-3mm}}

\newcommand{\Author}[1]{\begin{flushleft}
                       \it #1 \end{flushleft}}

\newcommand{\Address}[1]{\begin{flushleft}
                       \it #1 \end{flushleft}}

\newcommand{\Date}[1]{\begin{flushleft}
                      \small  \it #1 \end{flushleft}}

\newcommand{\evenhead}{Author \ name}
\newcommand{\oddhead}{Article \ name}

\renewcommand{\@evenhead}{
\hspace*{-3pt}\raisebox{-15pt}[\headheight][0pt]{\vbox{\hbox to \textwidth
{\thepage \hfil \evenhead}\vskip4pt \hrule}}}
\renewcommand{\@oddhead}{
\hspace*{-3pt}\raisebox{-15pt}[\headheight][0pt]{\vbox{\hbox to \textwidth
{\oddhead \hfil \thepage}\vskip4pt\hrule}}}
\renewcommand{\@evenfoot}{}
\renewcommand{\@oddfoot}{}

\long\def\@makecaption#1#2{%
  \vskip\abovecaptionskip
  \sbox\@tempboxa{\small \textbf{#1.}\ \ #2}%
  \ifdim \wd\@tempboxa >\hsize
    {\small \textbf{#1.}\ \ #2}\par
  \else
    \global \@minipagefalse
    \hb@xt@\hsize{\hfil\box\@tempboxa\hfil}%
  \fi
  \vskip\belowcaptionskip}

\newcommand{\JNMPnumberwithin}[3][\arabic]{%
  \@ifundefined{c@#2}{\@nocounterr{#2}}{%
    \@ifundefined{c@#3}{\@nocnterr{#3}}{%
      \@addtoreset{#2}{#3}%
      \@xp\xdef\csname the#2\endcsname{%
        \@xp\@nx\csname the#3\endcsname .\@nx#1{#2}}}}%
}

\newcommand{\resetfootnoterule} {
  \renewcommand\footnoterule{%
  \kern-3\p@
  \hrule\@width.4\columnwidth
  \kern2.6\p@}
}

\renewcommand{\footnoterule}{}

\newcommand{\be}{\begin{equation}}
\newcommand{\ee}{\end{equation}}
\newcommand{\ba}{\hspace*{-5pt}\begin{array}}
\newcommand{\ea}{\end{array}}
\newcommand{\p}{\partial}

\makeatother
\numberwithin{equation}{section}
\theoremstyle{definition}

\renewcommand{\ba}{\begin{array}}
\renewcommand{\ea}{\end{array}}
\newcommand{\beg}{\begin{eqnarray}}
\newcommand{\eeq}{\end{eqnarray}}
\newcommand{\bg}{\begin{eqnarray*}}

\newcommand{\ed}{\end{eqnarray*}}

\newcommand{\nn}{\nonumber}

\renewcommand{\p}{\partial} 
 
\newcommand{\notlhd}{\lhd\kern-.8em{/}\ } 
\newcommand{\notexist}{\ \exists\kern-.5em{\raise.1em\hbox{/}}\ }

\newcommand{\pde}[2]{\frac{\p #1}{\p #2}} 
 
\newcommand{\pdd}[2]{\frac{\p^2 #1}{\p #2^2}} 
\newcommand{\inp}{{\mbox{\vbox{\hrule width0ex\hbox{\vrule
 height0ex\kern3.8pt
\vbox{\kern2.5pt}\kern3.8pt \vrule height1.6ex}
\hrule width1.6ex}}}}

\begin{document}

\renewcommand{\evenhead}{ {\LARGE\textcolor{blue!10!black!40!green}{{\sf \ \ \ ]ocnmp[}}}\strut\hfill 
M Euler and N Euler
}
\renewcommand{\oddhead}{ {\LARGE\textcolor{blue!10!black!40!green}{{\sf ]ocnmp[}}}\ \ \ \ \   
New 5th-order Schwarzian evolution equations 
}

%%%% Matter for the first page 
\thispagestyle{empty}
\newcommand{\FistPageHead}[3]{
\begin{flushleft}
\raisebox{8mm}[0pt][0pt]
{\footnotesize \sf
\parbox{150mm}{{Open Communications in Nonlinear Mathematical Physics}\ \  \ {\LARGE\textcolor{blue!10!black!40!green}{]ocnmp[}}
\ \ Vol.6 (2026) pp
#2\hfill {\sc #3}}}\vspace{-13mm}
\end{flushleft}}

\FistPageHead{1}{\pageref{firstpage}--\pageref{lastpage}}{ \ \ Article}

\strut\hfill

\strut\hfill

\copyrightnote{The author(s). Distributed under a Creative Commons Attribution 4.0 International License}

%\copyrightnote{The author(s). Distributed under a Creative Commons Attribution 4.0 International License}

%\renewcommand{\theequation}{\arabic{equation}}

\renewcommand{\theequation}{\arabic{section}.\arabic{equation}}

\allowdisplaybreaks

%\renewcommand{\evenhead}{ {\LARGE\textcolor{blue!10!black!40!green}{{\sf \ \ \ ]ocnmp[}}}\strut\hfill 
%O Ragnisco and F Zullo
%}
%\renewcommand{\oddhead}{ {\LARGE\textcolor{blue!10!black!40!green}{{\sf ]ocnmp[}}}\ \ \ \ \   
%The integrable Volterra system and infinitely many species
%}

%\renewcommand{\evenhead}{}
%\renewcommand{\oddhead}{}

% Title

\thispagestyle{empty}

\Name{
New 5th-order Schwarzian evolution equations and their higher-order symmetries{\footnote{The reviewing process for this article was handled by Giuseppe Gaeta.}}
}

\label{firstpage}

\strut\hfill

\Author{Marianna Euler and Norbert Euler}

\Address{
International Society of Nonlinear Mathematical Physics, Auf der Hardt 27,
56130 Bad Ems, Germany \& Centro Internacional de Ciencias, Av. Universidad s/n,
Colonia Chamilpa, 62210 Cuernavaca, Morelos, Mexico
}

\Date{Received August 18, 2026; Accepted November 12, 2026}

%\renewcommand{\theequation}{\thesection.\arabic{equation}}

%\renewcommand{\theequation}{\arabic{equation}}

%\strut\hfill

\setcounter{equation}{0}

\strut\hfill In memory of Wilhelm I. Fushchych.\qquad\,\,\,

\strut\hfill On the occasion of his 90th anniversary.

\strut\hfill

\smallskip

\noindent
{\bf Citation format for this Article:}\newline
 Marianna Euler and Norbert Euler, 
New 5th-order Schwarzian evolution equations and their higher-order symmetries,
{\it Open Commun. Nonlinear Math. Phys.}, {\bf 6}, ocnmp:19000, \pageref{firstpage}--\pageref{lastpage}, 2026.

\strut\hfill

\noindent
{\bf The permanent Digital Object Identifier (DOI) for this Article:}\newline
{\it 10.46298/ocnmp.19000}

\strut\hfill

\noindent
{\bf Abstract:} We report new quasilinear and fully-nonlinear 5th-order Schwarzian evolution equations. These are symmetry-integrable evolution equations in 1+1 dimensions, i.e. equations that admit Lie-Bäcklund symmetries, whereby it is required that the equations are kept invariant under the Möbius transformation for their dependent variable. 

\section{Introduction}

%Definition of a Schwarzian Equation (based on our 2019 article)

%In this section we establish the notation and recall the notion of the Schwarzian evolution equations and their corresponding $\tau$-equations. In Section 2 we summarize all known Schwarzian evolution equations of order three and order five and give their corresponding $\tau$-equations up to order seven explicitly. This is essential for a comparison with the new Schwarzian equations that are reported in Section 3 and Section 4.

%\subsection{Preliminaries: Schwarzian equations and their $\tau$-equations}
%First we recall some facts about Schwarzian evolution equations.
A {\it Schwarzian evolution equation} of order $n$, as introduced in \cite{E-E-76}, is of the form
\begin{gather}
\label{Sch-EE-n}
u_t=u_x\Phi(S,S_x,S_{xx},\ldots,S_{(n-3)x}),
\end{gather}
where $S$ is the Schwarzian derivative
\begin{gather}
\label{S-der}
S:=\frac{u_{3x}}{u_x}-\frac{3}{2}\frac{u_{xx}^2}{u_x^2}
\end{gather}
and the function $\Phi$ is such that equation (\ref{Sch-EE-n}) admits an infinite number of local commuting Lie-Bäcklund symmetries with Lie-Bäcklund symmetry generators of the form
\begin{gather}
\label{LB-SG}
Z^u=Q(u,u_x,\ldots,u_{px})\pde{\ }{u},\quad p>n.
\end{gather}
Here $Q$ defines the Lie-Bäcklund invariant surface of order $p$ for (\ref{Sch-EE-n}), which can be expressed in the form
\begin{gather}
Q=u_x\Psi(S,S_x,\ldots,S_{(p-3)x})
\end{gather}
for some function $\Psi$  \cite{E-E-76}.
%Here and throughout this paper we use the notation $V_{xx}=\p^2 V/\p x^2$ and $V_{px}=\p^pV/\p x^p$ for $p>2$. 

Note that the more general non-autonomous case, where $\Phi$ depends explicitly on $x$ and $t$, i.e.
%namely the non-autonomous Schwarzian evolution equations
\begin{gather}
u_t=u_x\Phi(x,t,S,S_x,\ldots,S_{(n-3)x}),
\end{gather}
will not be considered here.

A Lie-Bäcklund symmetry (also known as generalised symmetries \cite{Olver-book})
of (\ref{Sch-EE-n}) and its corresponding invariant surface naturally defines another Schwarzian evolution equation that we name the {\it $\tau$-equation} of order $p$, namely
\begin{gather}
\label{u-tau-EE-Gen}
u_\tau=u_x\Psi(S,S_x,\ldots,S_{(p-3)x}),
\end{gather}
whereby
\begin{gather}
D_t u_\tau=D_\tau u_t.
\end{gather}
It is well known that if one Lie-Bäcklund symmetry generator (\ref{LB-SG}) exists for a given scalar equation (\ref{Sch-EE-n}) then this equation admits infinitely many Lie-Bäcklund symmetry generators of higher order, which then identifies  (\ref{Sch-EE-n}) as a symmetry-integrable equation \cite{Fokas}. This generates a hierarchy of symmetry-integrable evolution equations that are all $\tau$-equations, whereby the base equation of the hierarchy is (\ref{Sch-EE-n}).
Furthermore, equation (\ref{Sch-EE-n}) is invariant under the projective transformation in $u$ (or Möbius transformation) and translations in $x$ and $t$, i.e.
%, that contain the real parameters $\alpha_j,\ \beta_j,\ \epsilon_j$, namely the transformation
\begin{subequations}
\begin{gather} 
u(x,t)\mapsto \frac{\alpha_1 u(x,t)+\beta_1}{\alpha_2u(x,t)+\beta_2},\quad \alpha_1\beta_2-\alpha_2\beta_1=1\\[0.3cm]
x\mapsto x+\epsilon_1,\qquad t\mapsto t+\epsilon_2.
\end{gather}
\end{subequations}
Here $\alpha_j,\ \beta_j$ and $\epsilon_j$ are real parameters. 

An important observation is that equation (\ref{Sch-EE-n}) can be expressed in terms of $S$ 
as follows \cite{E-E-76}: 
%the $n$th-order auxiliary $S$-equation (given by Lemma 1 in \cite{E-E-76}) as follows:
\begin{gather}
\label{S-eq-Gen}
S_t=(D_x^3+2SD_x+S_x)\Phi(S,S_x,\ldots,S_{(n-3)x}). 
\end{gather}
Equation (\ref{S-eq-Gen}) is known as the {\it auxiliary $S$-equation} to the Möbius-invariant equation 
(\ref{Sch-EE-n}).
Clearly (\ref{Sch-EE-n})  is symmetry-integrable if and only if (\ref{S-eq-Gen}) is symmetry-integrable. In particular, 
equation (\ref{S-eq-Gen}) admits a Lie-Bäcklund symmetry generator of order $p$
\begin{gather}
Z^S=\left(D_x^3+2SD_x+S_x\right)\Psi(S,S_x,\ldots,S_{(p-3)x})\pde{\ }{S}
\end{gather}
if and only if (\ref{Sch-EE-n}) admits a Lie-Bäcklund symmetry generator of order $p$
\begin{gather}
Z^u=u_x\Psi(S,S_x,\ldots,S_{(p-3)x})\pde{\ }{u},
\end{gather}
whereby the corresponding $\tau$-equation in terms of $S$ is the auxiliary $S$-equation of (\ref{u-tau-EE-Gen}), i.e.
\begin{gather}
\label{Sch-EE-S-p}
S_\tau=\left(D_x^3+2SD_x+S_x\right)\Psi(S,S_x,\ldots,S_{(p-3)x}),
\end{gather}
with
\begin{gather}
D_\tau S_t=D_t S_\tau.
\end{gather}
%On the other hand, (\ref{Sch-EE-S-p}) is merely the auxiliary $S$-equation of (\ref{u-tau-EE-Gen}).

Note further that the auxiliary $S$-equation
(\ref{S-eq-Gen}) is never fully-nonlinear (meaning nonlinear in its highest derivative $S_{px}$), and therefore it is usually easier to analyse (\ref{S-eq-Gen}) rather than  (\ref{Sch-EE-n}).

\strut\hfill

In the current paper we are interested in the 5th-order case. That is, we aim to find all Schwarzian equations of the form
\begin{gather}
\label{Sch-EE-5th}
u_t=u_x\Phi(S,S_x,S_{xx})
\end{gather}
with its auxiliary $S$-equation
\begin{gather}
\label{Aux-S-EE-5th}
S_t=(D_x^3+2SD_x+S_x)\Phi(S,S_x,S_{xx}). 
\end{gather}
%This can be achieved by establishing the Lie-Bäcklund symmetries of the auxiliary $S$-equation of (\ref{Sch-EE-5th}), namely
%\begin{gather}
%\label{Aux-S-EE-5th}
%S_t=(D_x^3+2SD_x+S_x)\Phi(S,S_x,S_{xx}). 
%\end{gather}
%
%\strut\hfill
We do so by finding all functions $\Phi$ such that (\ref{Sch-EE-5th}) admits a Lie-Bäcklund symmetry generator $Z^u$ of order seven, i.e. 
\begin{gather}
Z^u=Q(u,u_x,\ldots,u_{7x})\pde{\ }{u}.
\end{gather}
%The auxiliary $S$-equation of (\ref{Sch-EE-5th}) is then
%\begin{gather}
%\label{Aux-S-EE-5th}
%S_t=(D_x^3+2SD_x+S_x)\Phi(S,S_x,S_{xx}). 
%\end{gather}
Here $Q$ must satisfy the symmetry invariance condition
\begin{gather}
\label{IC}
\left.
\vphantom{\frac{DA}{DB}}
L_E[u]Q\right|_{E=0}=0,
\end{gather}
where $E:=u_t-u_x\Phi(S,S_x,S_{xx})$ and
$L_E[u]$ denotes the linear operator
\begin{subequations}
\begin{gather}
L[u]=\pde{E}{u_t}D_t+\pde{E}{u}
+\pde{E}{u_x}D_x
+\pde{E}{u_{xx}}D_x^2
+\cdots+
\pde{E}{u_{5x}}D_x^5.
\end{gather}
\end{subequations}
The function $Q$ can be expressed in the form
\begin{gather}
Q(u,u_x,\ldots,u_{7x})=u_x\Psi(S,S_x,\ldots,S_{4x})
\end{gather}
and corresponds to the 7th-order $\tau$-equation
\begin{gather}
\label{Sch-EE-u-7th}
u_\tau=u_x\Psi(S,S_x,\ldots,S_{4x})
\end{gather}
with its auxiliary $S$-equation 
\begin{gather}
\label{Sch-EE-S-7th}
S_\tau=\left(D_x^3+2SD_x+S_x\right)\Psi(S,S_x,\ldots,S_{4x})
\end{gather}
given by 7th-order Lie-Bäcklund symmetry generator
\begin{gather}
Z^S=\left(D_x^3+2SD_x+S_x\right)\Psi(S,S_x,\ldots,S_{4x})\pde{\ }{S}.
\end{gather}

\smallskip

The paper is organized as follows:
In Section 2 we summarize all known Schwarzian evolution equations of order three and order five and give their corresponding $\tau$-equations up to order seven explicitly. 
%This is essential for a comparison with the new Schwarzian equations that are reported in Section 3 and Section 4.
In Section 3 we report all 5th-order quasilinear Schwarzian equations and in Section 4 all fully-nonlinear 5th-order Schwarzian equations. We identify the new Schwarzian equations by comparing our results to those reported earlier in \cite{E-E-76} and \cite{EE-JNMP-2020}. In each case, we give the corresponding 7th-order $\tau$-equations (\ref{Sch-EE-u-7th}) explicitly. In Section 5 we make some concluding remarks.

\smallskip

We should point out that the semilinear 3rd-order Schwarzian equation and all semilinear 5th-order Schwarzian equations, as well as all fully-nonlinear 3rd-order Schwarzian equations, have previously been reported in \cite{E-E-76}. In this sense, the current paper can be seen as a continuation of \cite{E-E-76}, whereby we now complete the 5th-order case by establishing all quasilinear 5th-order and all fully-nonlinear 5th-order Schwarzian equations. These new Schwarzian equations are not $\tau$-equations associated with 3rd-order Schwarzian equations. To make this clear and to make the comparison easier, we summarize in the next Section the Schwarzian equations of order three and order five that were previously obtained and reported in \cite{E-E-76} and include all corresponding $\tau$-equations up to order seven.

\section{The known Schwarzian evolution equations of orders three and five}

\subsection{The 3rd-order semilinear Schwarzian equation and its semilinear $\tau$-equations up to order seven}

A 3rd-order semilinear Schwarzian evolution equation and its auxiliary $S$-equation must be of the form
\begin{subequations}
\begin{gather}
\label{SKdV-u-gen}
u_t=u_x(\lambda_1 S+\lambda_2),\\[0.3cm]
\label{SKdV-S-gen}
S_t=\left(D_x^3+2SD_x+S_x\right)(\lambda_1 S+\lambda_2),
\end{gather}
\end{subequations}
where $\lambda_1$ and $\lambda_2$ are real constants with $\lambda_1\neq 0$. Without any loss of generality we can set $\lambda_1=1$ and $\lambda_2=0$, by which (\ref{SKdV-u-gen}) is the well-known Schwarzian Korteweg-de Vries equation and (\ref{SKdV-S-gen}) the Korteweg-de Vries equation, i.e.
\begin{subequations}
\begin{gather}
\label{SKdV-u}
u_t=u_xS,\\[0.3cm]
\label{SKdV-S}
S_t=S_{3x}+3SS_x,
\end{gather}
\end{subequations}
respectively. The corresponding 5th-order $\tau_1$-equations are
\begin{subequations}
\begin{gather}
\label{SKdV-u-tau-1}
u_{\tau_1}=u_x\left(S_{xx}+\frac{3}{2}S^2\right)\\[0.3cm]
\label{SKdV-S-tau-1}
S_{\tau_1}=S_{5x}+5SS_{3x}
+10 S_xS_{xx}+\frac{15}{2}S^2S_x,
\end{gather}
\end{subequations}
and the 7th-order $\tau_2$-equations are
\begin{subequations}
\begin{gather}
\label{SKdV-u-tau-2}
u_{\tau_2}=u_x\left( S_{4x}+5SS_{xx}
+\frac{5}{2}S_x^2+\frac{5}{2}S^3 \right)\\[0.3cm]
\label{SKdV-S-tau-2}
S_{\tau_2}=S_{7x}
+7SS_{5x}
+21S_xS_{4x}
+\frac{35}{2}\left(2S_{xx}+S^2\right)S_{3x}
+70SS_xS_{xx}\nn\\[0.3cm]
\qquad
+\frac{35}{2}S_x^3
+\frac{35}{2}S^3S_x,
\end{gather}
\end{subequations} 
where 
\begin{gather}
D_{\tau_i}u_{\tau_j}=D_{\tau_j}u_{\tau_i},\
D_{\tau_i}S_{\tau_j}=D_{\tau_j}S_{\tau_i},\ 
%\\[0.3cm] 
\mbox{for all}\ i>j,\  i,j\in\{0,1,2\},\ \tau_0\equiv t.
\end{gather}
It is well-known that (\ref{SKdV-u}) admits a 2nd-order recursion operator (given, for example, in 
\cite{Euler-book-2018}). The $\tau_3$-equation in $u$ is a semilinear Schwarzian equation of order nine and can easily be calculated using the mentioned recursion operator, and the same for the higher-order $\tau$-equations.

\subsection{The 5th-order semilinear Schwarzian equations and their semilinear $\tau$-equations up to order seven}

A 5th-order semilinear Schwarzian evolution equation and its auxiliary S-equation must be
 of the form
\begin{subequations}
\begin{gather}
\label{Schw-5th-Semi-gen}
u_t=u_xS_{xx}+u_xF(S,S_x),\\[0.3cm]
\label{Schw-S-5th-Gen}
S_t=S_{5x}+2SS_{3x}+S_xS_{xx}
+\left(D_x^3+2SD_x+S_x\right)F(S,S_x),
\end{gather}
\end{subequations}
respectively. Here the explicit form of the function $F$ is determined by the Lie-Bäcklund symmetry invariance condition (\ref{IC}). This has led  to two semilinear 5th-order Schwarzian equations, which we list below together with their auxiliary $S$-equations as reported in \cite{E-E-76}. We also include here the corresponding 7th-order $\tau_1$-equations.

\strut\hfill

\noindent
{\bf Case 1.3.1:}
The following semilinear 5th-order Schwarzian equation and its auxiliary $S$-equation was reported in \cite{E-E-76}:
\begin{subequations}
\begin{gather}
\label{KS-I-u}
u_t=u_x\left(S_{xx}+\frac{1}{4}S^2\right)\\[0.3cm]
\label{KS-I-S}
S_t=S_{5x}+\frac{5}{2}SS_{3x}
+\frac{5}{2}S_xS_{xx}+\frac{5}{4}S^2S_x.
\end{gather}
\end{subequations}
The corresponding 7th-order $\tau_1$-equations are 
\begin{subequations}
 \begin{gather}
 u_{\tau_1}=u_x\left(
 S_{4x}+\frac{3}{2}SS_{xx}
+\frac{3}{4}S_x^2+\frac{1}{6}S^3\right)\\[0.3cm]
S_{\tau_1}=S_{7x}
+\frac{7}{2}SS_{5x}
+7S_xS_{4x}
+\frac{7}{2}\left(S^2+3S_{xx}\right)S_{3x}
+\frac{7}{8}S^3S_x\nn\\[0.3cm]
\qquad
+\frac{21}{2}SS_xS_{xx}
+\frac{7}{4}S_x^3.
\end{gather}
\end{subequations}
Note that a 6th-order recursion operator $R[u]$ for (\ref{KS-I-u}) is reported in 
\cite{E-E-76}, where it was shown that there exist two hierarchies of semilinear Schwarzian equations with the base equation (\ref{KS-I-u}): one hierarchy is generated by $R[u]^ju_x$, so the $\tau_2$-equation for this hierarchy is of order seven, the $\tau_3$ equation of order thirteen, etc.. The second hierarchy is generated by $R[u]^j u_t$, whereby  the corresponding $\tau_2$ equation is of order eleven and the $\tau_3$-equation of order seventeen, etc. It is important to note that there exists here no $\tau$-equation of order nine as in the case of the  Schwarzian Korteweg-de Vries hierarchy.

\strut\hfill

\noindent
{\bf Case 1.3.2:}
The following semilinear 5th-order Schwarzian equation and its auxiliary $S$-equation was reported in \cite{E-E-76}:
\begin{subequations}
\begin{gather}
\label{KS-II-u}
u_t=u_x\left(S_{xx}+4S^2\right)\\[0.3cm]
\label{KS-II-S}
S_t=S_{5x}+10SS_{3x}
+25 S_xS_{xx}+20S^2S_x.
\end{gather}
\end{subequations}
The corresponding 7th-order $\tau_1$-equations are
\begin{subequations}
 \begin{gather}
 u_{\tau_1}=u_x\left(
 S_{4x}+12SS_{xx}
+6S_x^2+\frac{32}{3}S^3\right)\\[0.3cm]
S_{\tau_1}=S_{7x}
+14SS_{5x}
+49 S_xS_{4x}
+56\left(
\frac{3}{2}S_{xx}+S^2\right)S_{3x}
+\frac{224}{3}S^3S_x\nn\\[0.3cm]
\qquad
+252SS_xS_{xx}
+70S_x^3.
\end{gather}
\end{subequations}
Note that, similar to the Case 1.3.1, the semilinear equation (\ref{KS-II-u}) also admits a 6th-order recursion 
operator $R[u]$ that generates two hierarchies as reported in \cite{E-E-76}.

\smallskip

In \cite{E-E-76} we have established that there exist, besides equations (\ref{KS-I-u}) and (\ref{KS-II-u}),
no other semilinear 5th-order Schwarzian evolution equations that are not $\tau$-equations of the Korteweg-de Vries hierarchy.

\subsection{The 3rd-order fully-nonlinear Schwarzian equations and their quasilinear $\tau$-equations up to order seven}

A 3rd-order fully-nonlinear Schwarzian evolution equation and its auxiliary $S$-equation must be of the form
\begin{subequations}
\begin{gather}
\label{3rd-order-FN-Gen-1}
u_t=u_x\Phi(S)\\[0.3cm]
\label{3rd-order-FN-Gen-2}
S_t=\left(D_x^3+2SD_x+S_x\right)\Phi(S),
\end{gather}
where
\begin{gather}
\frac{d^2\Phi}{d S^2}\neq 0.
\end{gather}
\end{subequations}
Corresponding to (\ref{3rd-order-FN-Gen-1}) -- (\ref{3rd-order-FN-Gen-2}), the 5th-order quasilinear $\tau_1$-equations
are of the form
\begin{subequations}
\begin{gather}
\label{5th-order-fromFN3-Gen-1}
u_{\tau_1}=u_x\Psi_1(S,S_x,S_{xx})\\[0.3cm]
\label{5th-order-fromFN3-Gen-2}
S_{\tau_1}=\left(D_x^3+2SD_x+S_x\right)\Psi_1(S,S_x,S_{xx}),
\end{gather}
\end{subequations}
and the 7th-order quasilinear $\tau_2$-equations
are of the form
\begin{subequations}
\begin{gather}
\label{7th-order-fromFN3-Gen-1}
u_{\tau_2}=u_x\Psi_2(S,S_x,\ldots,S_{4x})\\[0.3cm]
\label{7th-order-fromFN3-Gen-2}
S_{\tau_2}=\left(D_x^3+2SD_x+S_x\right)\Psi_2(S,S_x,\ldots,S_{4x}).
\end{gather}
\end{subequations}
%%%%%%%%%%%%%%%%%%%%%%%%%%
Following this notation, we now list the three essential fully-nonlinear 3rd-order Schwarzian equations reported in \cite{E-E-76}, as well as their 5th-order 
$\tau_1$-equations which were derived and reported in
\cite{EE-JNMP-2020}. For completeness we include the 7th-order $\tau_2$-equations for each case.

\strut\hfill

\noindent
{\bf Case 1.4.1:} Consider the fully-nonlinear 3rd-order Schwarzian equation of the form (\ref{3rd-order-FN-Gen-1}) 
with its auxiliary $S$-equation (\ref{3rd-order-FN-Gen-2}), where  
\begin{gather}
\label{Phi-Case141}
\Phi(S)=-2S^{-1/2}.
\end{gather}
This fully-nonlinear 3rd-order equation, viz.
\begin{gather}
\label{Case141-u-3rd}
u_t=-2u_xS^{-1/2}
\end{gather}
was established in \cite{E-E-76}.
The 5th-order quasilinear $\tau_1$-equations are then given by (\ref{5th-order-fromFN3-Gen-1}) -- 
(\ref{5th-order-fromFN3-Gen-2}),
where
\begin{gather}
\Psi_1(S,S_x,S_{xx})=
S^{-5/2}S_{xx}-\frac{5}{4}S^{-7/2}S_x^2+4S^{-1/2}
\end{gather}
as given in \cite{EE-JNMP-2020}.
The 7th-order quasilinear $\tau_2$-equations are given by (\ref{7th-order-fromFN3-Gen-1}) -- (\ref{7th-order-fromFN3-Gen-2}),
where
\begin{gather}
%\label{Psi_2-Case-141}
\Psi_2(S,S_x,\ldots,S_{4x})=
S^{-7/2}S_{4x}
-7S^{-9/2}S_xS_{3x}
-\frac{21}{4}S^{-9/2}S_{xx}^2
-2S^{-5/2}S_{xx}\nn\\[0.3cm]
\label{Psi_2-Case-141}
\qquad
+\frac{231}{8}S^{-11/2}S_x^2S_{xx}
-\frac{1155}{64}S^{-13/2}S_x^4
+\frac{5}{2}S^{-7/2}S_x^2
-8S^{-1/2}.
\end{gather}
Note that the fully-nonlinear equation (\ref{Case141-u-3rd}) admits a 2nd-order recursion operator that 
was obtained earlier and reported in \cite{E-E-O-2025}. The $\tau_3$-equation 
is therefore a quasilinear equation of order nine and can easily be calculated using this recursion operator.

\strut\hfill

\noindent
{\bf Case 1.4.2:} Consider the fully-nonlinear 3rd-order Schwarzian equation of the form (\ref{3rd-order-FN-Gen-1}) 
and its auxiliary $S$-equation (\ref{3rd-order-FN-Gen-2}), where
\begin{gather}
\label{Phi-Case142}
\Phi(S; b_1)=(b_1-S)^{-2}
\end{gather}
with $b_1$ an arbitrary real constant. This fully-nonlinear 3rd-order equation, viz.
\begin{gather}
\label{Case142-u-3rd}
u_t=u_x(b_1-S)^{-2}
\end{gather}
was established in \cite{E-E-76}.
The 5th-order quasilinear $\tau_1$-equations are then given by (\ref{5th-order-fromFN3-Gen-1}) -- 
(\ref{5th-order-fromFN3-Gen-2}),
where
\begin{gather}
\Psi_1(S,S_x,S_{xx}; b_1)=\frac{4S_{xx}}{(b_1-S)^{5}}
+\frac{10S_x^2}{(b_1-S)^{6}}
-\frac{b_1-4S}{(b_1-S)^{4}}
\end{gather}
as given in \cite{EE-JNMP-2020}.
The 7th-order quasilinear $\tau_2$-equations are given by (\ref{7th-order-fromFN3-Gen-1}) -- (\ref{7th-order-fromFN3-Gen-2}),
where
\begin{gather}
\Psi_2(S,S_x,\ldots,S_{4x};b_1)=\frac{S_{4x}}{(b_1-S)^7}
+\frac{14S_xS_{3x}}{(b_1-S)^8}
+\frac{21S_{xx}^2}{2(b_1-S)^8}
+\frac{5SS_{xx}}{(b_1-S)^7}\nn\\[0.3cm]
\qquad
+\frac{98S_x^2S_{xx}}{(b_1-S)^9}
+\frac{189S_x^4}{2(b_1-S)^{10}}
+\frac{5(b_1+6S)S_x^2}{2(b_1-S)^8}
+\frac{15S^2-6b_1S+b_1^2}{8(b_1-S)^6}.
\end{gather}
Note that the fully-nonlinear equation (\ref{Case142-u-3rd}) admits a 2nd-order recursion operator 
that was obtained earlier and reported in  \cite{EE-OCNMP-2022}. The $\tau_3$-equation 
is therefore a quasilinear equation of order nine and can easily be calculated using this recursion operator.

\strut\hfill

\noindent
{\bf Case 1.4.3:} Consider the fully-nonlinear 3rd-order Schwarzian equation of the form (\ref{3rd-order-FN-Gen-1}) 
and its auxiliary $S$-equation (\ref{3rd-order-FN-Gen-2}), where
\begin{gather}
\Phi(S;a_1,a_2)=\frac{a_1-S}{(a_1^2+3a_2)(S^2-2a_1 S-3a_2)^{1/2}}.
\end{gather}
Here $a_1$ and $a_2$ are real constants such that $a_1^2+3a_2\neq 0$. 
This fully-nonlinear 3rd-order equation, viz.
\begin{gather}
\label{Case143-u-3rd}
u_t=\frac{u_x(a_1-S)}{(a_1^2+3a_2)(S^2-2a_1 S-3a_2)^{1/2}}
\end{gather}
was established in \cite{E-E-76}.

The 5th-order quasilinear $\tau_1$-equations are then given by (\ref{5th-order-fromFN3-Gen-1}) -- 
(\ref{5th-order-fromFN3-Gen-2}),
where
\begin{gather}
%\label{Case143-Phi1}
\Psi_1(S,S_x,S_{xx};a_1,a_2)=
\frac{S_{xx}}{(S^2-2a_1 S-3a_2)^{5/2}}
+\frac{5(a_1-S)S_x^2}{(S^2-2a_1 S-3a_2)^{7/2}}\nn\\[0.3cm]
\label{Case143-Phi1}
\qquad
+\frac{2a_1S^3
-6a_1^2S^2
+3a_1(a_1^2-3a_2)S
+3a_2(a_1^2-3a_2)}
{(a_1^2+3a_2)^2(S^2-2a_1S-3a_2)^{3/2}}
\end{gather}
as given in \cite{EE-JNMP-2020} (note that the expression for $\Psi_1$ in \cite{EE-JNMP-2020} contains a misprint).
The 7th-order quasilinear $\tau_2$-equations are given by (\ref{7th-order-fromFN3-Gen-1}) -- (\ref{7th-order-fromFN3-Gen-2}),
where
\begin{gather}
\Psi_2(S,S_x,\ldots,S_{4x};a_1,a_2)=
\frac{S_{4x}}{(S^2-2a_1S-3a_2)^{7/2}}
+\frac{14(a_1-S)S_xS_{3x}}
{(S^2-2a_1S-3a_2)^{9/2}}\nn\\[0.3cm]
\qquad
+\frac{21(a_1-S)S_{xx}^2}
{(S^2-2a_1S-3a_2)^{9/2}}
+\frac{7(28S^2-56a_1S+33a_1^2+15a_2)S_x^2S_{xx}}
{2(S^2-2a_1S-3a_2)^{11/2}}\nn\\[0.3cm]
\qquad
+\frac{5S_xS_{xx}}
{(S^2-2a_1S-3a_2)^{7/2}}
+\frac{21(a_1-S)(36S^2-72a_1S+55a_1^2+57a_2)S_x^4}
{8(S^2-2a_1S-3a_2)^{13/2}}\nn\\[0.3cm]
\qquad
-\frac{5(6S^2-5a_1S+3a_2)S_x^2}
{2(S^2-2a_1S-3a_2)^{9/2}}
-\frac{3(a_1^2-a_2)S^5}
{(a_1^2+3a_2)^3
(S^2-2a_1S-3a_2)^{5/2}}\nn\\[0.3cm]
\qquad
+\frac{15a_1(a_1^2-a_2)S^4}
{(a_1^2+3a_2)^3
(S^2-2a_1S-3a_2)^{5/2}}
+\frac{45(a_1^2-a_2)^2S^3}
{2(a_1^2+3a_2)^3
(S^2-2a_1S-3a_2)^{5/2}}\nn\\[0.3cm]
\qquad
+\frac{15a_1(a_1-9a_2)(a_1^2-a_2)S^2}
{2(a_1^2+3a_2)^3
(S^2-2a_1S-3a_2)^{5/2}}
+\frac{15a_1^2a_2(a_1^2-9a_2)S}
{(a_1^2+3a_2)^3
(S^2-2a_1S-3a_2)^{5/2}}\nn\\[0.3cm]
\qquad
+\frac{9a_1a_2^2(a_1^2-9a_2)}
{(a_1^2+3a_2)^3
(S^2-2a_1S-3a_2)^{5/2}}.
\end{gather}
We have so far not been able to find a recursion operator for the fully-nonlinear equation (\ref{Case143-u-3rd}). However, we have established and reported a 6th-order recursion operator for the $S$-equation corresponding to (\ref{Case143-u-3rd}) in \cite{EE-JNMP-2020}, where it is shown how to use the recursion operators of the quasilinear $S$-equations to generate the hierarchy of the fully-nonlinear equations in $u$.

\section{Quasilinear 5th-order Schwarzian equations}
A 5th-order quasilinear Schwarzian evolution equation must be of the form
\begin{gather}
\label{Quasi-5th-u-Gen}
u_t=u_x\left[\Phi_1(S,S_x)S_{xx}+\Phi_2(S,S_x)\right],
\end{gather}
whereby its auxiliary $S$-equation is also quasilinear and of the form
\begin{gather}
\label{Quasi-5th-S-Gen}
S_t=\left(D_x^3+2SD_x+S_x\right)\left[\Phi_1(S,S_x)S_{xx}+\Phi_2(S,S_x)\right].
\end{gather}
We now establish the explicit forms of $\Phi_1$ and $\Phi_2$ such that (\ref{Quasi-5th-u-Gen}) admits a 7th-order
Lie-Bäcklund symmetry using the invariance condition (\ref{IC}), whereby we will exclude the 5th-order quasilinear 
$\tau$-equations listed in Section 1.4. The resulting 5th-order Schwarzian equations that we obtained in this way are given by the following two cases:

\strut\hfill

\noindent
{\bf Case 2.1:} Consider equation (\ref{Quasi-5th-u-Gen}), viz,
\begin{gather*}
%\label{Quasi-5th-u-Gen}
u_t=u_x\left[\Phi_1(S,S_x)S_{xx}+\Phi_2(S,S_x)\right]
\end{gather*}
with
\begin{gather}
\label{Quasi-5th-u-Gen-Cond}
\pde{\Phi_1}{S_x}\neq 0.
\end{gather}
This leads to 

\begin{proposition}
The only 5th-order Schwarzian evolution equations (\ref{Sch-EE-5th}), viz.
%\label{Aux-S-EE-5th}
%
\begin{gather*}
u_t=u_x\Phi(S,S_x,S_{xx}),
\end{gather*}
for which  (\ref{Sch-EE-5th}) is of the form (\ref{Quasi-5th-u-Gen}) under the condition
(\ref{Quasi-5th-u-Gen-Cond}) and not a $\tau$-equation associated with the fully-nonliner 3rd-order Schwarzian equations, is given by the following functional form of $\Phi$:
\begin{gather}
%\label{Case-1-Phi}
\Phi(S,S_x,S_{xx};\alpha,\beta)=\left(S_x+\alpha S^3+\frac{\beta}{\alpha}\right)^{-5/3}S_{xx}\nn\\[0.3cm]
\label{Case-1-Phi}
\quad\ 
-3\alpha S^2\left(S_x+\alpha S^3+\frac{\beta}{\alpha}\right)^{-5/3}\left(\alpha S^3+\frac{\beta}{\alpha}\right)
+\frac{15\alpha S^2}{2}\left(S_x+\alpha S^3+\frac{\beta}{\alpha}\right)^{-2/3}.
\end{gather}
Here $\alpha$ is an arbitrary but non-zero constant and $\displaystyle{\beta\in\{\frac{1}{18},\ \frac{8}{9}\}}$.
The auxiliary $S$-equation is of the form (\ref{Aux-S-EE-5th})
%, viz.
%\begin{gather*}
%\label{Aux-S-EE-5th}
%S_t=(D_x^3+2SD_x+S_x)\Phi(S,S_x,S_{xx};\alpha,\beta). 
%\end{gather*}
with $\Phi$ given by (\ref{Case-1-Phi}). The corresponding 
7th-order $\tau_1$-equations are (\ref{Sch-EE-u-7th}) and 
(\ref{Sch-EE-S-7th}), viz.
% and its 7th-order auxiliary $S$-equation are of the form (\ref{Sch-EE-u-7th}) and 
%(\ref{Sch-EE-S-7th}), respectively, viz.
\begin{gather*}
u_{\tau_1}=u_x\Psi(S,S_x,\ldots,S_{4x};\alpha,\beta)\\[0.3cm]
S_{\tau_1}=(D_x^3+2SD_x+S_x)\Psi(S,S_x,\ldots,S_{4x};\alpha,\beta),
\end{gather*}
where
\begin{gather}
\Psi(S,S_x,\ldots,S_{4x};\alpha,\beta)=
\left(S_x+\alpha S^3+\frac{\beta}{\alpha}\right)^{-7/3}S_{4x}\nn\\[0.3cm]
\quad\ 
-\frac{14}{3}\left(S_x+\alpha S^3+\frac{\beta}{\alpha}\right)^{-10/3}S_{xx}S_{3x}
-14\alpha \left(S_x+\alpha S^3+\frac{\beta}{\alpha}\right)^{-10/3}S^2S_xS_{3x}\nn\\[0.3cm]
\quad\ 
+\left(S_x+\alpha S^3+\frac{\beta}{\alpha}\right)^{-13/3}
\left[\frac{35}{9}
S_{xx}^3
+\frac{7}{2}\left(
7\alpha 
S^2S_{x}
-3\alpha^2S^5
-3\beta S^2\right)S_{xx}^2
\right.\nn\\[0.3cm]
\quad\
-7\alpha SS_x^3S_{xx}
+\frac{1}{5}\left(
490 \alpha^2S^3
+133\beta
+4\right)SS_x^2S_{xx}
+\frac{(231\beta+8)(\alpha^2S^3+\beta)}{5\alpha}
SS_xS_{xx}\nn\\[0.3cm]
\quad\ 
+\frac{(63\beta +4)(\alpha^2 S^3+\beta)^2}{5\alpha^2}
SS_{xx}
+\frac{21\alpha}{2}S_x^5
-\frac{3(315\alpha^2S^3-49\beta +8)}{10}S_x^4\nn\\[0.3cm]
\quad\ 
-\frac{3\alpha^2(231\beta+8)S^3+15\beta(7\beta+3)}{5\alpha}S_x^3
- \frac{39\beta(\alpha^2 S^3+\beta)^2(9\beta +2)}{10\alpha^3}S_x
\nn\\[0.3cm]
\quad\ 
-\frac{3(\alpha^2S^3+\beta)(4\alpha^2S^3+63\alpha^2\beta S^3+21\beta+87\beta^2)}{5\alpha^2}S_x^2
\nn\\[0.3cm]
\left.
\quad\ 
-\frac{9\beta(\alpha^2S^3+\beta)^3(9\beta+2)}{10\alpha^4}\right].
\end{gather} 
%with $\alpha$ an arbitrary non-zero constant and $\displaystyle{\beta\in\{\frac{1}{18},\,\frac{8}{9}\}}$

\end{proposition}

%%%%
%%%%

\strut\hfill

\noindent
{\bf Case 2.2:} Consider the quasilinear equation 
\begin{gather}
\label{Quasi-5th-u-Gen-2}
u_t=u_x\left[\Phi_1(S)S_{xx}+\Phi_2(S,S_x)\right],
\end{gather}
whereby its auxiliary $S$-equation is also quasilinear and of the form
\begin{gather}
\label{Quasi-5th-S-Gen}
S_t=\left(D_x^3+2SD_x+S_x\right)\left[\Phi_1(S)S_{xx}+\Phi_2(S,S_x)\right].
\end{gather}
This leads to 

\begin{proposition}
The only 5th-order Schwarzian evolution equations (\ref{Sch-EE-5th}), viz.
%\label{Aux-S-EE-5th}
%
\begin{gather*}
u_t=u_x\Phi(S,S_x,S_{xx}),
\end{gather*}
for which (\ref{Sch-EE-5th}) is quasilinear of the form (\ref{Quasi-5th-u-Gen-2}) 
and not a $\tau$-equation associated with the fully-nonlinear Schwarzian 3rd-order equations, is given by the following functional form of $\Phi$:
\begin{gather}
%\label{Case-2-Phi}
\Phi(S,S_x,S_{xx};\alpha)=(S+\alpha)^{-5/3}S_{xx}
-\frac{5}{3}(S+\alpha)^{-8/3}S_x^2\nn\\[0.3cm]
\label{Case-2-Phi}
\quad\ 
+\frac{3}{2}(S+\alpha)^{-2/3}\left(S+\frac{3\alpha}{2}\right),
\end{gather}
where $\alpha$ is an arbitrary constant.
The auxiliary $S$-equation is of the form (\ref{Aux-S-EE-5th}), viz.
\begin{gather*}
%\label{Aux-S-EE-5th}
S_t=(D_x^3+2SD_x+S_x)\Phi(S,S_x,S_{xx};\alpha)
\end{gather*}
with $\Phi$ given by (\ref{Case-2-Phi}). 
The corresponding 
7th-order $\tau_1$-equations are (\ref{Sch-EE-u-7th}) and 
(\ref{Sch-EE-S-7th}), viz.
\begin{gather*}
u_{\tau_1}=u_x\Psi(S,S_x,\ldots,S_{4x};\alpha)\\[0.3cm]
S_{\tau_1}=(D_x^3+2SD_x+S_x)\Psi(S,S_x,\ldots,S_{4x};\alpha),
\end{gather*}
where
\begin{gather}
\Psi(S,S_x,\ldots,S_{4x};\alpha)=
(S+\alpha)^{-7/3}S_{4x}
-7(S+\alpha)^{-10/3}S_xS_{3x}\nn\\[0.3cm]
\quad\ 
+\frac{245}{9}(S+\alpha)^{-13/3}S_x^2S_{xx}
-\frac{14}{3}(S+\alpha)^{-10/3}S_{xx}^2
+\frac{3}{2}(S+\alpha)^{-7/3}SS_{xx}\nn\\[0.3cm]
\quad\ 
-\frac{455}{27}(S+\alpha)^{-16/3}S_x^4
-\frac{1}{12}(S+\alpha)^{-10/3}(26S-9\alpha)S_x^2\nn\\[0.3cm]
\quad\ 
+\frac{3}{8}(S+\alpha)^{-4/3}\left(2S^2+12\alpha S+9\alpha^2\right).
\end{gather} 
%with $\alpha$ an arbitrary constant.

\end{proposition}

\section{Fully-nonlinear 5th-order Schwarzian equations}

Consider the fully-nonlinear 5th-order equation 
\begin{subequations}
\begin{gather}
\label{Quasi-5th-u-Gen-FN-a}
u_t=u_x\Phi(S,S_x,S_{xx})\\[0.3cm]
\label{Quasi-5th-u-Gen-FN-b}
\mbox{with}\ \pdd{\Phi}{S_{xx}}\neq 0.
\end{gather}
\end{subequations}
The corresponding auxiliary $S$-equation is quasilinear and of the form
\begin{gather}
\label{Quasi-5th-S-Gen}
S_t=\left(D_x^3+2SD_x+S_x\right)\Phi(S,S_x,S_{xx}).
\end{gather}
This leads to the following

\begin{proposition}
The only 5th-order fully-nonlinear Schwarzian evolution equations are given by
(\ref{Quasi-5th-u-Gen-FN-a}), whereby $\Phi$ takes any of the following three functional forms:
%%%
\begin{subequations}
\begin{gather}
\label{5th-SI-Phi-j}
\Phi_j(S,S_x,S_{xx};\alpha_j)=S^{5/6}\left(S_{xx}-\frac{5}{4}\frac{S_x^2}{S}+\alpha_j S^2\right)^{-2/3},\quad j=1,2;\\[0.3cm]
\label{5th-SI-Phi-3}
\Phi_3(S,S_x,S_{xx})=S^{10/9}\left(S_{xx}-\frac{5}{3}\frac{S_x^2}{S}+\frac{3}{2}S^2\right)^{-2/3}.
\end{gather}
\end{subequations}
Here
\begin{gather}
\label{alpha-j}
\alpha_1=\frac{2}{3}\ \ \mbox{and}\  \ \alpha_2=-\frac{8}{3}.
\end{gather}
The auxiliary $S$-equations are of the form (\ref{Aux-S-EE-5th}), viz.
\begin{gather*}
%\label{Aux-S-EE-5th}
S_t=(D_x^3+2SD_x+S_x)\Phi_k(S,S_x,S_{xx};\alpha_j),\quad k=1,2,3
\end{gather*}
where $\Phi_k$ are given by (\ref{5th-SI-Phi-j}) --  (\ref{5th-SI-Phi-3}). The corresponding 
7th-order $\tau_1$-equations are of the form 
(\ref{Sch-EE-u-7th}) and 
(\ref{Sch-EE-S-7th}), viz.
\begin{gather*}
u_{\tau_1}=u_x\Psi_k(S,S_x,\ldots,S_{4x};\alpha_j)\\[0.3cm]
S_{\tau_1}=(D_x^3+2SD_x+S_x)\Psi_k(S,S_x,\ldots,S_{4x};\alpha_j),
\end{gather*}
where
\begin{subequations}
\begin{gather}
\Psi_j(S,S_x,\ldots,S_{4x};\alpha_j)=S^{7/6}S_{4x}\left(
S_{xx}-\frac{5}{4}\frac{S_x^2}{S}+\alpha_j S^2\right)^{-7/3}\nn\\[0.3cm]
\quad\ 
-S^{-17/6}\left(
S_{xx}-\frac{5}{4}\frac{S_x^2}{S}+\alpha_j S^2\right)^{-10/3}
\left(
7\alpha_j S^5S_xS_{3x}
+\frac{7}{6}S^4S_{3x}^2
\right.\nn\\[0.3cm]
\quad\ 
-\frac{7}{2}S^3S_xS_{xx}S_{3x}
+\frac{1365}{64}\alpha_j S^3S_x^4
+\frac{15}{16}S^3S_x^4
-\frac{69}{16} \alpha_j^2S^6S_x^2
-\frac{3}{2}\alpha_j S^6S_x^2\nn\\[0.3cm]
\quad\ 
-\frac{1575}{256}S_x^6
+\frac{27}{20}\alpha_j^3S^9
+\frac{3}{5}\alpha_j^2S^9
+\frac{69}{20}\alpha_j^2S^7S_{xx}
+\frac{6}{5}\alpha_j S^7S_{xx}\nn\\[0.3cm]
\quad
-\frac{273}{8}\alpha_j S^4S_x^2S_{xx}
-\frac{3}{2}S^4S_x^2S_{xx}^2
+\frac{945}{64}SS_x^4S_{xx}
+\frac{147}{20}\alpha_j S^5S_{xx}^2\nn\\[0.3cm]
\quad\ 
\label{Phi-S-Case2}
\left.
+\frac{3}{5}S^5S_{xx}^2
-\frac{189}{16}S^2S_x^2S_{xx}^2
+\frac{21}{4}S^3S_{xx}^3\right),\quad j=1,2
\end{gather}
with $\alpha_j$ given by (\ref{alpha-j}), and
\begin{gather}
\Psi_3(S,S_x,\ldots,S_{4x})=S^{14/9}S_{4x}
\left(
S_{xx}-\frac{5}{3}\frac{S_x^2}{S}+\frac{3}{2}S^2
\right)^{-7/3}\nn\\[0.3cm]
\quad\ 
-S^{-22/9}\left(
S_{xx}-\frac{5}{3}\frac{S_x^2}{S}+\frac{3}{2}S^2
\right)^{-10/3}
\left(
\frac{7}{6}
S^4S_{3x}^2
+\frac{35}{3}
S^5S_xS_{3x}
-\frac{35}{27}S^2S_x^3S_{3x}\right.\nn\\[0.3cm]
\quad\ 
-\frac{14}{3}S^3S_xS_{xx}S_{3x}
+7S^3S_{xx}^3
-\frac{35}{2} S^2S_x^2S_{xx}^2
+16S^5S_{xx}^2
+\frac{51}{4}S^7 S_{xx} \nn\\[0.3cm]
\quad\ 
\left.
-77 S^4S_x^2 S_{xx} 
+\frac{805}{27}S S_x^4S_{xx}
-\frac{515}{24}S^6S_x^2
-\frac{3850}{243} S_x^6
+\frac{3025}{54}S^3S_x^4
\right.
\nn\\[0.3cm]
\label{Phi-S-Case1}
\quad\ 
\left.
+\frac{27}{4}S^9\right).
\end{gather}
\end{subequations}
\end{proposition}

\noindent
We remark that case (\ref{5th-SI-Phi-j}) was derived earlier and reported in \cite{E-E-O-2025} as part of the classification of symmetry-integrable evolution equations invariant under the following projective transformation
\begin{subequations}
\begin{gather} 
u(x,t)\mapsto \frac{\alpha_1 u(x,t)+\beta_1}{\alpha_2u(x,t)+\beta_2},\quad \alpha_1\beta_2-\alpha_2\beta_1=1
\\[0.3cm]
x\mapsto \frac{\gamma_1 x+\delta_1}{\gamma_2 x+\delta_2},\quad \gamma_1\delta_2-\delta_2\gamma_1=1\\[0.3cm]
t\mapsto t+\epsilon,
\end{gather}
\end{subequations}
where $\alpha_j,\ \beta_j,\ \gamma_j$ and $\epsilon$ are real parameters.

\section{Concluding remarks}
We have identified several new 5th-order Schwarzian evolution equations by establishing the 7th-order Lie-Bäcklund symmetries of the general Möbius-invariant 5th-order equation 
\begin{gather*}
u_t=u_x\Phi(S,S_x,S_{xx}),
\end{gather*}
where $S$ is the Schwarzian derivative (\ref{S-der}). As far as we know, these equations have not been reported before. For the readers' convenience we list below those Schwarzian 5th-order equations explicitly:
\begin{subequations}
\begin{gather}
%\label{Con-1}
u_t=u_x\left[\left(S_x+\alpha S^3+\frac{1}{18\alpha}\right)^{-5/3}S_{xx}
-3\alpha S^2\left(S_x+\alpha S^3+\frac{1}{18\alpha}\right)^{-5/3}\left(\alpha S^3+\frac{1}{18\alpha}\right)
\right.\nn\\[0.3cm]
\label{Con-1}
\quad\ 
\left.
+\frac{15\alpha S^2}{2}\left(S_x+\alpha S^3+\frac{1}{18\alpha}\right)^{-2/3}\right],\nn\\
\label{Con-10}
\quad\ 
\mbox{where\ $\alpha$ is an arbitrary but non-zero constant};\\[0.3cm]
%
%
%
%\label{Con-10}
u_t=u_x\left[\left(S_x+\alpha S^3+\frac{8}{9\alpha}\right)^{-5/3}S_{xx}
-3\alpha S^2\left(S_x+\alpha S^3+\frac{8}{9\alpha}\right)^{-5/3}\left(\alpha S^3+\frac{8}{9\alpha}\right)
\right.
\nn\\[0.3cm]
\quad\ 
\left.
+\frac{15\alpha S^2}{2}\left(S_x+\alpha S^3+\frac{8}{9\alpha}\right)^{-2/3}\right],\nn\\
%\label{Con-10}
\quad\ 
\mbox{where\ $\alpha$ is an arbitrary but non-zero constant};\\[0.3cm]
u_t=u_x\left[(S+\alpha)^{-5/3}S_{xx}
-\frac{5}{3}(S+\alpha)^{-8/3}S_x^2
+\frac{3}{2}(S+\alpha)^{-2/3}\left(S+\frac{3\alpha}{2}\right)\right],\nn\\
\quad\ 
\mbox{where $\alpha$ is an arbitrary constant};\\[0.3cm]
u_t=
u_xS^{5/6}\left(S_{xx}-\frac{5}{4}\frac{S_x^2}{S}+\frac{2}{3} S^2\right)^{-2/3};\\[0.3cm]
u_t=
u_xS^{5/6}\left(S_{xx}-\frac{5}{4}\frac{S_x^2}{S}-\frac{8}{3} S^2\right)^{-2/3};\\[0.3cm]
\label{Con-2}
u_t=u_xS^{10/9}\left(S_{xx}-\frac{5}{3}\frac{S_x^2}{S}+\frac{3}{2}S^2\right)^{-2/3}.
\end{gather}
\end{subequations}
We emphasis that none of the equation (\ref{Con-10}) -- (\ref{Con-2}) admit a Lie-Bäcklund symmetry of order nine.
Instead,  the orders of their Lie-Bäcklund symmetries are $7,11,13,17,19,...$ That is, the symmetries of of orders 1 (mod 6) and 5 (mod 6). At this point we have not yet established the recursion operators for these new Schwarzian equations. As mentioned before, a recursion operator for the 3rd-order fully-nonlinear equation (\ref{Case143-u-3rd}) given in Case 1.4.3 is also outstanding.

As far as we know, there are currently no 7th-order Schwarzian evolution equations known that are not $\tau$-equations of 3rd-order or 5th-order Schwarzian equations. This could be an interesting future project.

It would furthermore be of interest to do a systematic study of all the possible potentialisations and to apply possible generalised hodograph transformations for the Schwarzian equations that are reported here, similar to the study in \cite{EE-OCNMP-2024} and \cite{E-E-2025-SI-Bluman}, where some fully-nonlinear symmetry-integrable equations were mapped to semilinear and quasilinear equations. This work is currently in progress.

\section*{Regarding the use of Artificial Intelligence (AI)}
The authors have not used AI tools to write any of the text or to perform any of the calculations that are reported here. 

\section*{Acknowledgements}
We are grateful to Ricardo Buring and Arthemy Kiselev for their help in solving a nonlinear algebraic system of equations which was necessary for the results stated in Proposition 3. We also thank Patrizia Rogolino and Francesco Oliveri for inviting and hosting us in Messina for the {\it MAGMAV 2026 Workshop} held during 12--14 January 2026, and to Francesco Oliveri for interesting discussions in Messina.

\label{lastpage}

\end{document}